\documentclass[preprint,aps]{revtex4} 
\usepackage{graphicx}

\begin{document}
\title{Fundamental spatiotemporal decoherence: \\a key to solving 
the conceptual
problems of black holes, cosmology and quantum mechanics}

\author{Rodolfo Gambini}
\address{Instituto de F\'{\i}sica, Facultad de Ciencias, 
Universidad de la Rep\'ublica, Igu\'a 4225, CP 11400 Montevideo,
Uruguay\\rgambini@fisica.edu.uy}
\author{Rafael A. Porto}
\address{Department of Physics, Carnegie Mellon University,
Pittsburgh, PA 15213\\rporto@andrew.cmu.edu}
\author{Jorge Pullin}
\address{Department of Physics and Astronomy, 
Louisiana State University, Baton Rouge, LA 70803-4001\\pullin@lsu.edu}

\date{March 28th 2006}

\begin{abstract}
Unitarity is a pillar of quantum theory. Nevertheless, it is also a
source of several of its conceptual problems. We note that in a world
where measurements are relational, as is the case in gravitation,
quantum mechanics exhibits a fundamental level of loss of
coherence. This can be the key to solving, among others, the puzzles
posed by the black hole information paradox, the formation of
inhomogeneities in cosmology and the measurement problem in quantum
mechanics.
\end{abstract}

\maketitle

In ordinary quantum field theory space-time is taken to be a classical
variable that can be measured with arbitrarily high precision. We know
however, that this is only an approximation. In reality there exist
fundamental limitations to how accurately we can measure space and
time. In recent papers we have argued that the lack of accuracy in
measuring time leads to a loss of unitarity in ordinary quantum
mechanics \cite{gapopudeco}. This loss of unitarity was
sufficiently significant to render the black hole information puzzle
unobservable \cite{gapopuprl}.  In this essay we would like to explore
the implications of the fact that space cannot be measured with
arbitrary precision either.

Pioneering work in this area was done by Salecker and Wigner
\cite{SaWi}, further elaborated by Ng and Van Dam
\cite{Ng} who noted that there are fundamental limitations to the accuracy
with which space-time can be measured.  It should be noted that these
limitations are due to the measurement apparatus and they tend to be
considerably larger than the intrinsic uncertainties due to the
quantum fluctuations of the metric. That is, even if we ignore the
latter and assume space-time to be classical, there will still be
limitations in the accuracy with which the former can be
measured. Covariantly, for a space-time interval $s$ the bound they
find for the error of measurement is $\delta |s| \sim L_{\rm Planck}
\sqrt[3]{|s|/L_{\rm Planck}}$. 

We will assume that space-time is flat for the rest of this paper,
though it is possible to generalize the results with the usual caveats
of quantum field theory on curved space-time. In spite of this, we are
entering into the discussion of quantum field theory crucial elements
that have to be used in formulating a theory of quantum
gravity. Namely, in quantum gravity (at least formulated canonically)
one starts with a spatial manifold with coordinates $r$, that is due
to a slicing of a space-time in which a time parameter $t$ has been
identified. Both the spatial and temporal variables $t,r$ are fiducial
variables in the construction and do not have physical meaning. To
give physical meaning to space-time points one has to recourse to
values of physical variables. For instance dust and scalar fields have
been used to give physical meaning to the coordinates \cite{BrKu}, see
also
\cite{GiHaMa}.

We will consider a set of fields $\phi(\vec{r},t)$ and we will assume
that one can, in the spirit of \cite{BrKu}, construct observables of
the theory whose level sets can be used to label physical points
$\vec{R}(\phi(\vec{r},t))$ and $T(\phi(\vec{r},t))$ (from now on we
will drop the vector symbol on the $r$ to simplify notation).  Quantum
mechanically, one would like to ask questions like ``what is the
probability that an observable of the field $\psi$ take a certain
value $\psi_0$ at a given coordinate point $\vec{R},T$ in
space-time''. Such question will be formulated as a conditional
probability, following the same lines as in the quantum mechanical
case \cite{gapopudeco}.

To make contact with ordinary quantum field theory we will assume a
semi-classical behavior for the clock and measuring rods in order to
compare with the ideal case. Given the similarities with previously
obtained results in the context of quantum mechanics \cite{gapopudeco}
we will directly jump into the final stages of the calculation.  We
assume the quantum state and evolution operators factorize into a
``clock and rods'' and a system under study.  We also assume the
density matrix for the system under study is given by $\rho_{sys}$.
One ends up with an effective density matrix as a function of the
clock and rod variables,
\begin{equation}
\rho(T,R) \equiv \int_{-\infty}^\infty dt dr U_{\rm sys}(t,r) \rho_{\rm sys}
 U_{\rm sys}(t,r)^\dagger {\cal P}_{t,r}(T,R),
\end{equation}
with ${\cal P}_{t,r}(T,R)$ the probability that the variables $T,R$
occur as a function of $t,r$, and from now on we drop the suffix
``sys'' in the density matrix. Again we refer the reader to the
quantum mechanical case for details \cite{njp}.

We have therefore ended with the standard probability expression with
an ``effective'' density matrix in the Schr\"odinger picture given by
$\rho(T,R)$. In this unorthodox representation the space-time dependence
of one field operator is shifted to the density matrix. We do this to 
keep things as parallel as possible to the quantum mechanical case, but
one later pays a price at the time of computing correlation functions, 
which are the usual observables of quantum field theory.  Now
that we have identified what will play the role of a density matrix in
terms of  ``real clocks and rods'' , we would like to see what
happens if we assume they are behaving semi-classically.  We start by
assuming that the probability for a measurement of $R,T$ for a given
value of $r,t$ is given by a function ${\cal P}_{t,r}(T,R)$ centered
at $t,r$ and with a width $b(s)$ where $(T,R)^\mu=s n^\mu$ and $n^\mu$
is a unitary four dimensional vector (we exclude in a first analysis
the use of null intervals for simplicity). Under these assumptions the
Lorentz invariant equation for the density matrix turns out to be,
\begin{equation}
{\partial \rho(s)\over \partial s} =i [\rho,P] +\sigma(s) 
[P,[P,\rho]]+...
\end{equation}
with $P\equiv P^\mu n_\mu$ and $P^\mu$ is the spacetime canonical
momentum operator.  The extra term is determined by the rate of
variation of the width of the distribution, $\sigma(s)=\partial
b(s)/\partial s$. The equation is function only of the invariant
space-time interval from the point $R,T$ at which the field operator
was originally defined to the origin selected for the measurement of
these ``quasi-Lorentzian'' coordinates.  A similar equation was
recently considered by Milburn \cite{Mi}, though he did not derive it
from the relational measurement of space time points. In his approach
he assumed $\sigma$ to be a constant, whereas for us it follows from
the fundamental uncertainty relation $\delta s= L_{\rm
  Planck}^{2/3}s^{1/3}$, which yields $b(s)=L_{\rm Planck}^{4/3}
(s_{\rm max}^{2/3}-(s_{\rm max}-s)^{2/3})$ where $s_{\rm max}$ is the
spatiotemporal ``length'' of the experiment. This leads to a
decoherence effect in the four-momentum basis,
\begin{equation}
  \rho(s)_{k_n,k_m} = \rho_{k_n,k_m}(0)
\exp\left({-i s(k^\mu_{n}-k^\mu_{m})n_\mu}\right) \exp\left({-\left[(k^\mu_{n}-k^\mu_{m})n_\mu\right]^2 L_{\rm Planck}^{4/3} s^{2/3}}\right).\label{p}
\end{equation} 
where we have chosen the ``optimal'' measurement device, i.e. $s_{\rm
max}=s$.  This is the spacetime extension of our previous result that
includes temporal decoherence as a particular case $n^i=0,n^0=1,s=T$.
We have presented results in terms of the density matrix.  We have
not completed an analysis in terms of the more familiar observables
of quantum field theory, correlation functions, but preliminary
calculations show a similar dependence with distance and the
difference of momentum components.

Is there a chance of observing experimentally these effects? Simon and
Jaksch \cite{SiJa} have argued that there is a better chance of
detecting spatial decoherence than the purely temporal one we had
proposed. The purely temporal decoherence needed the construction of
quantum state superpositions with energy differences between levels of
$10^{10}eV$. Though not out of the question in the future, such states
are not available in the lab today. For the spatial decoherence
introduced here, better experiments could be conceived.

Now that we have established the basis for spatio-temporal decoherence,
let us remark briefly on the possible implications of this mechanism
for some of the most basic conceptual problems of modern physics:

$\bullet$ The construction we have carried out is the most natural to
deal with space and time in a diffeomorphism invariant theory, leading
to a relational description of nature that solves ``the problem of
time'' in a well defined implementation of canonical quantum gravity
where the conditional probabilities can be computed without conceptual
obstructions and which originated this point of view
\cite{greece,ashtekarws}. As such, it appears as natural and
conceptually strong, yet it leads to a fresh perspective on several
conceptual problems in physics.

$\bullet$ As we argued in \cite{gapopuprl} having a fundamental mechanism of
loss of coherence is able to render the black hole information puzzle
unobservable.  The fact that pure states evolve naturally into mixed
states implies that there is no puzzle in such an evolution when it
occurs when a black hole evaporates.  In addition to this a detailed
calculation \cite{gapopuprl} shows that the order of magnitude of the speed of
the effect is appropriate in the context of black hole evaporation.

$\bullet$ The mechanism for fundamental decoherence can lead to a
better understanding of the measurement problem in quantum mechanics.
In the usual system-environment interaction the off-diagonal terms of
the density matrix oscillate as a function of time. Since the
environment is usually considered to contain a very large number of
degrees of freedom, the common period of oscillation for the
off-diagonal terms to recover non-vanishing values is very large, in
many cases larger than the life of the universe. This allows to
consider the problem solved in practical terms, yet one is left with a
conceptual puzzle: could the off diagonal terms at least in principle
reappear?. When one adds the effect we discussed, since it
suppresses exponentially the off-diagonal terms, one never has the
possibility that the latter will see their initial values
restored, no matter how long one waits.

$\bullet$ In the standard inflationary paradigm for cosmology it is
assumed that quantum fluctuations in the inflaton give rise to the
cosmological perturbations that seed the formation of structure in the
universe. A puzzle arises since the fluctuations in the inflaton are
quantum in nature whereas the seeds for structure formation are
classical. Several mechanisms related to environmental decoherence
have been proposed to solve the puzzle. A recent critique
\cite{PeSaSu} however argues strongly that new physics is needed to
fully explain the quantum to classical transition. Our mechanism can
provide the needed ``new physics'' required for such a transition.
Given its momentum dependence, our effect would naturally destroy
correlations of short wavelength. Recalling that the environmental
decoherence makes the reduced density matrix for the large modes
obtained by tracing out the unobservable short wavelength modes, take
a quasi diagonal form, our effect would allow to explain the passage
from this reduced matrix to a true statistical mixture of long
wavelength inhomogeneities that will seed the formation of structure.

Summarizing, the use of the natural relational ideas needed to discuss
the physics of gravity yields modifications in quantum theory. Though
too small to be observed in the lab today, the modifications are 
profound enough to alter our understanding of several of the 
most challenging conceptual problems of modern physics.

This work was supported in part by grants NSF-PHY-0244335,
NSF-PHY-0554793, DOE-ER-40682-143 and DEAC02-6CH03000, and by funds of
the Horace C. Hearne Jr. Institute for Theoretical Physics, PEDECIBA
(Uruguay), FQXi and CCT-LSU.


\begin{references}
\bibitem{gapopudeco}
See R.~Gambini, R.~Porto and J.~Pullin,
  arXiv:gr-qc/0603090 and references therein.
\bibitem{gapopuprl} 
R.~Gambini, R.~A.~Porto and J.~Pullin,
  Phys.\ Rev.\ Lett.\  {\bf 93}, 240401 (2004)
  [arXiv:hep-th/0406260].
\bibitem{SaWi} E. Wigner, Rev. Mod. Phys. {\bf 29}, 255 (1957);
H. Salecker, E. Wigner, Phys. Rev. {\bf 109}, 571 (1958).
\bibitem{Ng}
Y.~J.~Ng and H.~van Dam,
Annals N.\ Y.\ Acad.\ Sci.\  {\bf 755}, 579 (1995) 
[arXiv:hep-th/9406110];
Mod.\ Phys.\ Lett.\ A {\bf 9}, 335 (1994).
\bibitem{BrKu}
J.~D.~Brown and K.~V.~Kuchar,
  Phys.\ Rev.\ D {\bf 51}, 5600 (1995)
  [arXiv:gr-qc/9409001]; 
 J.~D.~Brown and D.~Marolf,
  Phys.\ Rev.\ D {\bf 53}, 1835 (1996)
  [arXiv:gr-qc/9509026].
\bibitem{GiHaMa} 
  S.~B.~Giddings, D.~Marolf and J.~B.~Hartle,
Phys.\ Rev.\ D {\bf 74}, 064018 (2006)
  [arXiv:hep-th/0512200].
\bibitem{njp} R. Gambini, R. Porto, J. Pullin,
New J.\ Phys.\  {\bf 6}, 45 (2004)
[arXiv:gr-qc/0402118].
\bibitem{Mi}
  G.~J.~Milburn, New. J. Phys. {\bf 8}, 96 (2006)
[arXiv:gr-qc/0308021].
\bibitem{SiJa} C. Simon, D. Jaksch Phys. Rev. {\bf A70}, 052104 (2004). 
\bibitem{greece}
R. Gambini, R.A. Porto and J. Pullin, 
In ``Recent developments in gravity'', K. Kokkotas, N. Stergioulas, editors,
World Scientific, Singapore (2003) [arXiv:gr-qc/0302064].
\bibitem{ashtekarws}
R.~Gambini and J.~Pullin,
  ``Discrete space-time,'' in ``100 Years of Relativity, Space Time Structure:
Einstein and Beyond'', Abhay Ashtekar (editor), World Scientific, Singapore
(2005) [arXiv:gr-qc/0505023].
\bibitem{PeSaSu}
  A.~Perez, H.~Sahlmann and D.~Sudarsky,
Class. Quan. Grav. {\bf 23}, 2317 (2006) [arXiv:gr-qc/0508100].


\end{references}
\end{document}